\documentclass[twoside,compsoc]{IEEEtran}
\usepackage{makecell}
\usepackage{hyperref}
\usepackage{array}
\usepackage{graphicx,amssymb,amsmath}
\usepackage{multicol}
\usepackage[noadjust]{cite}
\usepackage{setspace}
\usepackage{subfigure}
\usepackage{float}
\usepackage {url}
\usepackage{stfloats}
\usepackage{amsthm,pifont}
\usepackage{flushend}
\usepackage{cases,subeqnarray}
\usepackage{bm,multirow,bigstrut}
\usepackage{amsmath, amsthm, amssymb}
\usepackage{textcomp}
\usepackage{latexsym,bm}
\usepackage{booktabs}
\usepackage{xcolor}
\usepackage{mathtools}
\usepackage{dsfont}
\usepackage{extarrows}
\usepackage{epsfig}
\usepackage{epsfig}
\usepackage{epstopdf}
\usepackage{colortbl}
\usepackage[noend]{algpseudocode}
\usepackage{algorithmicx,algorithm}
\IEEEoverridecommandlockouts
\begin{document}
%----------------title&author&thanks-------------
\title{Ubiquitous Intelligence Via  Wireless Network-Driven LLMs Evolution}
% :Opportunities and Challenges}
\author{Xingkun Yin, Feiran You, Hongyang Du, and Kaibin Huang
\thanks{X. Yin, F. You, H.~Du, and K.~Huang are with the Department of Electrical and Electronic Engineering, University of Hong Kong, Hong Kong SAR, China (email: yinxingkun@connect.hku.hk, youfr@hku.hk, duhy@hku.hk, huangkb@hku.hk).}
}
\maketitle

\vspace{-1cm}
%%%%%%%%%% Abstraction %%%%%%%%%%%%%
\begin{abstract}
We introduce \emph{ubiquitous intelligence} as a paradigm where Large Language Models (LLMs) evolve within wireless network-driven ecosystems. Unlike static model deployments, this approach enables scalable and continuous intelligence ascension through coordination between networks and LLMs. Wireless networks support system-orchestrated lifelong learning, while LLMs drive the next-generation network development that is more adaptive and responsive. This co-evolution highlights a shift toward self-improving systems, sustaining capability growth across diverse and resource-constrained environments.
\end{abstract}

%%%%%%%%%% Introduction %%%%%%%%%%%%%
\section{Introduction}
Large Language Models (LLMs) have quickly expanded from their original applications in machine translation and summarization to a wide spectrum of complex generation tasks, including code, graphics, and video~\cite{vaswani2023attentionneed}.
Advances such as ultra-long context processing, multimodal integration, and frameworks like Retrieval-Augmented Generation (RAG)~\cite{lewis2021retrievalaugmentedgenerationknowledgeintensivenlp} have further pushed LLMs into domains such as law and medicine traditionally relying on human expertise.
Over time, LLMs have evolved from passive decision-making assistants to active participants in end-to-end processes, while the emerging Artificial Intelligence (AI) agent~\cite{xi2023risepotentiallargelanguage} paradigm further extends its capabilities to autonomous reasoning, planning, and execution.

Despite these advances, most LLMs still operate as cloud-centric models, relying on large clusters for inference and periodic offline retraining~\cite{raza2025industrial}. This architecture delivers scale but struggles to meet the growing demands for low latency, strong privacy, and adaptive personalization. Wireless networks, which connect billions of heterogeneous edge devices, offer a promising alternative. By enabling distributed inference and continual learning closer to data sources, they provide the foundation for more responsive, private, and personalized LLMs deployment. Such a transition is technically feasible and increasingly necessary to sustain the next generation of large-scale intelligence.

\subsection{Large Language Models}
LLMs have emerged as a primary expression of machine intelligence, demonstrating the ability to generalize across diverse tasks. Their effectiveness relies on large-scale training and inference pipelines, which are predominantly deployed in centralized data centers~\cite{raza2025industrial, hu2024characterizationlargelanguagemodel}. These infrastructures integrate high-performance accelerators, low-latency interconnects, and deep memory hierarchies to support large-batch optimization and high-throughput inference~\cite{xu2024unleashing}. Leading commercial platforms, including OpenAI’s GPT models on Microsoft Azure GPU clusters~\cite{sanchez2024azure}, Google’s Gemini and PaLM on TPU-based infrastructures~\cite{alto2024building}, and DeepSeek’s multi-GPU distributed framework~\cite{deng2025exploring}, leverage this architecture for scalable, consistent training and inference. Services, such as NVIDIA’s DGX Cloud, further extend these capabilities via cloud-hosted multi-node inference, supporting enterprise-scale workloads~\cite{li2019evaluating}. This cloud-centric structure offers key advantages: on-demand scalability, centralized management, and streamlined maintenance, making it the standard for contemporary foundation models.

However, as LLMs become increasingly integrated into daily life, the limitations of the cloud-centric paradigm are growing more apparent. Development priorities now emphasize low latency, strong privacy, and adaptive personalization, yet transmitting data to and from remote data centers conflicts with the demands of time-sensitive applications~\cite{le2024applications,blika2024federated}. Effective personalization further requires continuous, context-aware learning from the user environment, but centralized processing of such data raises significant privacy risks~\cite{cooper2025advancingpersonalizedfederatedlearning}. These challenges highlight a widening gap between centralized infrastructures and user expectations, underscoring the need to shift LLM inference closer to data sources through wireless networks.

\subsection{Wireless Network}
Wireless networks form a crucial interface between cloud-centric AI and context-aware inference at the edge~\cite{le2024applications,zhang2024beyond,cox2022memory}. Modern infrastructures, characterized by dense connectivity, high throughput, and low latency, offer a flexible foundation for collaborative training and synchronized model updates across heterogeneous devices~\cite{hu2021distributed,khoramnejad2025generative,he2025integrating}. As LLMs are deployed in increasingly diverse scenarios, conventional reliance on centralized data centers imposes significant limitations. Transmission delays hinder responsiveness in time-sensitive applications~\cite{zhang2024edgeshardefficientllminference}, while personalized services require handling user-specific data previously ignored~\cite{liang2025largelanguagemodelswireless}. These demands exceed the capabilities of traditional network pipelines.

To meet these emerging challenges, wireless components such as Base Stations (BSs), access points, and user devices must evolve from passive relays into active participants in distributed reasoning~\cite{wen2024integrated,wu2023split,fang2023collaborative}. Through Device-to-Device (D2D) links, adaptive edge caching, and opportunistic spectrum access, network nodes can exchange intermediate computations, propagate refined models, and adapt decision policies in real time. This architectural transformation is central to the vision of Sixth-Generation (6G) networks~\cite{dang2020should}, which integrate communication, computation, and learning into a unified, human-centric system. By embedding intelligence directly within the network fabric, future systems aim to support personalized, secure, and ultra-reliable services under highly dynamic conditions.

%%%%%%%%% Scaling AI in the Wild %%%%%%%%%%%%%%
\section{Motivation}
LLMs and wireless networks have driven major advances in intelligence and connectivity, but both now face fundamental limits. Addressing these constraints is key to sustaining large-scale evolution.

\begin{figure}[t]
\centering
\includegraphics[width = 0.48\textwidth]{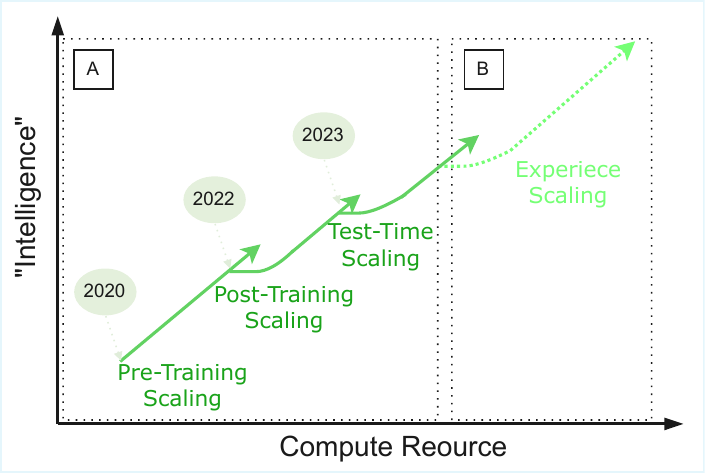}
\caption{Scaling Laws: Scaling in terms of parameter size, training data, and compute time can be conceptualized as scaling the amount of compute resources. LLMs have progressed through three stages of scaling laws, culminating in their current advanced stage. We propose a new scaling paradigm, experience scaling, within the concept of \emph{ubiquitous intelligence}, aiming to push scaling to a new dimension and further enhance LLMs' capabilities.}
\label{fig_new_scaling}
\end{figure}

\subsection{Developments and Challenges of Large Language Models}
The progression of LLMs has been driven by successive scaling paradigms (Fig.~\ref{fig_new_scaling}). Introduced in 2020, pre-training scaling law~\cite{kaplan2020scalinglawsneurallanguage}, established a power-law relationship between performance and three factors: \textbf{model size, training data, and compute}, guided the development of GPT-3~\cite{brown2020languagemodelsfewshotlearners} and later models. Post-training scaling focuses on improving alignment and task specialization through supervised fine-tuning and Reinforcement Learning from Human Feedback (RLHF)~\cite{chen2025p2lawscalinglaw,ouyang2022traininglanguagemodelsfollow}, allowing capability gains without changing the core architecture. More recently, test-time scaling emphasizes allocating additional compute during inference, enabling step-by-step reasoning, exploring multiple solution paths, and reducing hallucinations, thereby improving overall reliability and reasoning quality~\cite{wei2023chainofthoughtpromptingelicitsreasoning,zhang2025surveytesttimescalinglarge}.

However, current scaling laws exhibit diminishing returns: GPT-4’s greatly increased scale over GPT-3 yielded diminishing per-token gains, while training data demand now surpasses human generation, detrimental for post-training methods like RLHF~\cite{ouyang2022traininglanguagemodelsfollow}. As shown in Fig.~\ref{fig_new_scaling}.B, LLMs performance is nearing its ceiling under this paradigm, necessitating a new scaling dimension for further advancement.

\subsection{Evolution and Limitations of Wireless Networks}
Wireless networks have evolved from voice-only systems to intelligent infrastructures that support data-driven and mission-critical applications~\cite{sabourin2025strategic}. From 3G’s mobile internet to 5G’s broadband and massive connectivity, each generation has improved capacity and latency. Ongoing 6G research seeks to integrate sensing, communication, and intelligence into a unified architecture~\cite{khoramnejad2025generative,vyas20246g,zhang2024guest}. Key advances such as Ultra-Reliable Low-Latency Communication (URLLC)~\cite{ishtiaq2024edge}, D2D communication~\cite{9877925}, and context-aware content delivery~\cite{10288419} enhance responsiveness and localization, while adaptive spectrum and resource management~\cite{sanjalawe2025review} improve efficiency under dynamic conditions.

However, challenges persist as LLM-based services require low latency, privacy, and personalization, which cloud-based designs often ignores~\cite{panda2025scalable}. Transmitting data to remote servers introduces delay and privacy risks~\cite{syed2025artificial,pivoto2025comprehensive,xu2024large,souza2024maintenance}, limiting real-time applications. Meanwhile, networks themselves face growing complexity. Device heterogeneity, mobility, and spectrum variation reduce the effectiveness of centralized control, causing resource contention and unstable performance~\cite{wang2024understanding,sanjalawe2025review}. Scaling distributed AI workloads further amplifies these limitations. Importantly, these issues are not only bottlenecks for deploying LLMs but also opportunities for co-optimization. LLMs can contribute to the network’s optimization, enabling context-aware scheduling, semantic compression, and adaptive coordination. Fully unlocking this potential, however, requires architectural alignment and real-time integration between learning models and communication protocols.

\subsection{Motivation for Coevolution between LLMs and Wireless Network}
The advances and limitations of LLMs and wireless networks highlight the opportunity for mutual reinforcement. The co-evolution of LLMs with wireless networks involves leveraging edge computing and decentralized learning~\cite{boateng2025survey,zhang2024beyond,chen2025towards}, where models are deployed closer to users through local processing on edge devices or base stations, allowing LLMs to continuously learning without relying on centralized cloud servers.

This perspective motivates the vision of \textbf{\emph{ubiquitous intelligence} via AI-enabled networks and network-enabled AI}, where intelligence is embedded throughout the wireless network infrastructure and adaptively distributed across heterogeneous devices to support real-time, context-aware, and personalized services at scale.

\begin{figure}[t]
\centering
\includegraphics[width = 0.48\textwidth]{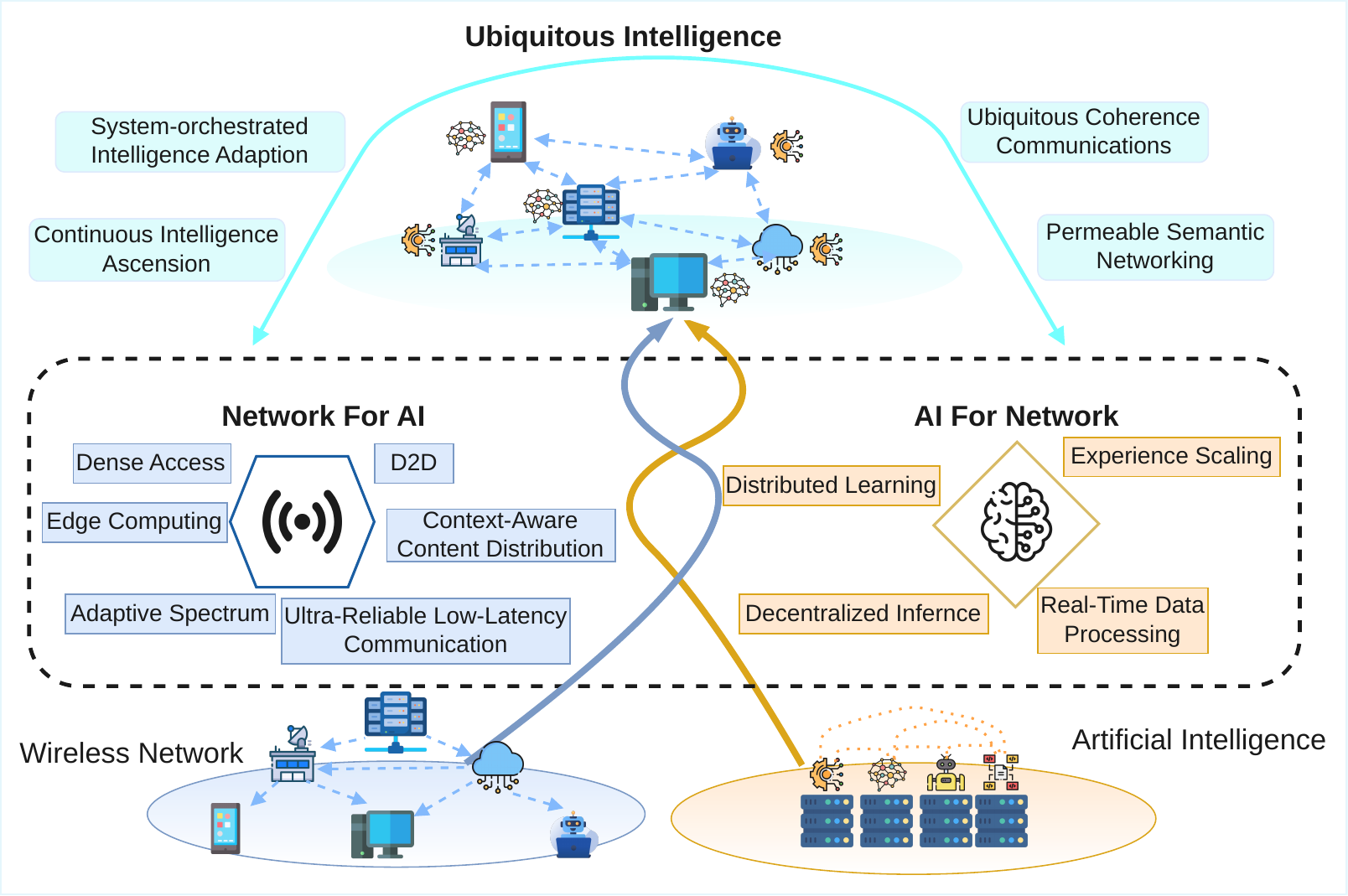}
\caption{Coevolution of Wireless Network and AI: Adaptive, continuously improving LLMs are supported through wireless network-assisted experience scaling and AI-powered, context-aware, densely connected networks, forming the foundation of \emph{ubiquitous intelligence}.
}
\label{fig_co-evolution}
\end{figure}

%%%%% Wireless as a Substrate %%%%%%%%%%%%%
\section{Synergistic Evolution of LLMs and Wireless Networks}
Achieving \emph{ubiquitous intelligence} requires tight integration between LLMs and network infrastructure. This section explores their co-evolution and mutual reinforcement in enabling pervasive cognition.

\subsection{Network for LLMs}
Although scaling laws have driven significant advances in LLMs, progress is limited as human-generated data can no longer keep pace with LLMs consumption~\cite{villalobos2024rundatalimitsllm}, signaling the need for a new paradigm. Current advances remain primarily focused on imitating humans, consuming human-created data. While LLMs are increasingly capable of performing existing tasks on behalf of humans, they still lack the capacity to deliver breakthroughs in scientific and technological domains where no human-generated data exists.

A new scaling paradigm is required to bypass the bottleneck of human-generated data by enabling LLMs to collect and create data directly from their environment. Advancements in LLMs and wireless networks (Fig.~\ref{fig_co-evolution}) enable autonomous knowledge acquisition through environmental interaction, extending model understanding beyond human-derived domains. This continuous, multimodal experience accumulation supports lifelong learning. When shared across wirelessly interconnected nodes, these heterogeneous data streams foster collective intelligence at scale, unlocking access to massive decentralized environmental data and paving the way for intelligence that transcends the limits of human experience.

\subsection{LLMs for Network}
% Framework (idea)
The advancement of LLMs now hinges not only on computational scaling but also on their integration with communication infrastructures. Wireless networks, once passive conduits for data, are increasingly envisioned as distributed platforms where models and network elements co-evolve in a tightly integrated ecosystem. This transformation enables ubiquitous intelligence, where computation and communication are jointly executed across heterogeneous, spatially distributed devices. The fusion of dense radio access, Multi-access Edge Computing (MEC), diverse terminals, and latency-aware scheduling provides a dynamic substrate for in-network learning~\cite{zeydan2025role}. Network nodes become intelligent agents, capable of contextual reasoning and localized model refinement. Devices actively participate in knowledge exchange and environmental adaptation, forming a collaborative learning system embedded within the communication fabric. Modern wireless architectures account for user behavior and localized demands~\cite{10288419,panda2025scalable,khoramnejad2025generative}, guiding the selective delivery of models or data to where they are most needed. This localized responsiveness is central to the realization of \emph{ubiquitous intelligence}.

\begin{figure}[t]
\centering
\includegraphics[width = 0.48\textwidth]{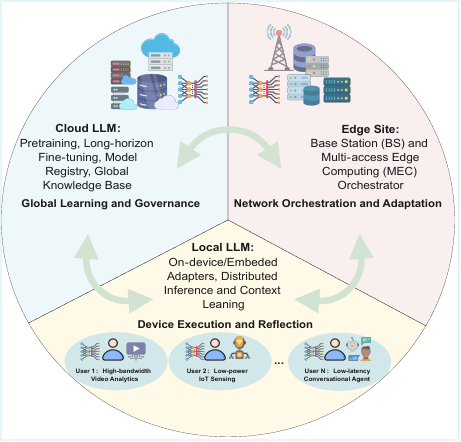}
\caption{\emph{Ubiquitous Intelligence}: LLMs evolve into adaptive, networked ecosystems integrated with wireless edge environments, enabling scalable, resilient, and continuously improving intelligent services.}
\label{fig_sys}
\end{figure}

Fig.~\ref{fig_sys} illustrates the framework of \emph{ubiquitous intelligence} with the deployment of cloud LLMs, edge sites, and local LLMs. Each layer participates in distributed inference, model refinement, and real-time collaboration. The close integration between radio access points and edge computation platforms~\cite{agarwal2025open} creates a tightly coupled infrastructure where models operate near the data source. This structure supports high availability, rapid adaptation, and robust performance under dynamic wireless conditions. As LLMs increasingly depend on the environments they inhabit, the wireless network becomes a partner in learning and reasoning. This shift marks a departure from isolated model scaling toward a co-evolutionary paradigm, where intelligence arises through continuous interaction between the models and the underlying wireless infrastructure.

\subsection{Ubiquitous Intelligence}
The trajectories of LLMs and wireless networks are increasingly convergent rather than independent. As demonstrated in Fig.~\ref{fig_co-evolution}, their integration marks a transition from incremental advances in each domain to a co-evolutionary paradigm in which intelligence and connectivity are inseparable. This convergence provides the conceptual and technical foundation for \textit{ubiquitous intelligence}, where adaptive learning and resilient communication form a unified, continuously evolving ecosystem. The shift towards \textit{ubiquitous intelligence} is grounded in four core principles:
\begin{itemize}
    \item {\textbf{Continuous Intelligence Ascension.}}    % Net4AI 外
    % experience scaling
    Continuous intelligence ascension denotes the sustained enhancement of LLMs capabilities through interaction with dynamic environments. Unlike static deployments, LLMs refine reasoning, adaptability, and autonomy from live experiences, while wireless networks interconnect heterogeneous sensing devices and edge nodes to provide the bandwidth and low latency needed for real-time experience exchange. This integration transforms learning into a distributed, lifelong process, enabling intelligence to scale continuously alongside wireless network development.
    
    \item {\textbf{System-Orchestrated Intelligence Adaption.}}
    % Net4AI 内
    % heterogeneous system different LLMs
    Aided by wireless network breakthroughs in D2D and context-aware content distribution, system-orchestrated intelligence adaptation supports heterogeneous LLMs frameworks to refine LLMs selectively by applying only the relevant subsets of experience according to each LLMs' size and function, thereby avoiding redundant updates and enhancing overall system efficiency. During inference, tasks are allocated to the most suitable LLMs based on task characteristics and resource availability, optimizing both performance and utilization. This principle embodies effective coordination and adaptive deployment of intelligence within distributed environments.
    
    \item {\textbf{Permeable Semantic Networking.}}
    % AI4Net 外
    %  LLM赋予的 semantic（message&infomation 的 semantic - 通信； services的semantic - 协同计算） 能力 带来网络系统能力的提升
    AI empowers networks with semantic awareness, allowing systems to exchange not only raw data but also the underlying meaning contained in messages and services. LLMs extend this semantic capability through their pre-trained alignment with human intent, enabling interpretation, generation, and abstraction of meaning across modalities. When integrated into communication and computation, LLMs enable semantic-level information exchange, transforming the network from a passive transport medium into an adaptive and context-aware infrastructure.

    \item {\textbf{Ubiquitous Coherence Communications.}}
    % AI4Net 内
    % 语言、推理和生成把混乱的信息空间逐步组织、压缩、提炼成有序知识; 从 “混乱 → 有序”、“冗余 → 精炼”、“扩散 → 收敛”
    LLM's reasoning and generation convert disordered information flows into structured and refined knowledge. Networks under this paradigm evolve from chaotic to organized, from redundant to essential, and from diffuse to convergent, thereby sustaining coherence in distributed communication environments. The integration of these capabilities ensures that ubiquitous connectivity is matched with ubiquitous intelligence, establishing networks as adaptive ecosystems for resilient, large-scale intelligent services.
    
\end{itemize}

\section{New Opportunities}
As LLMs evolve from static inference engines to dynamic cognitive systems, new opportunities emerge for continual refinement beyond initial training. This section examin
es how interaction, feedback, and decentralized adaptation within wireless networks support this ongoing evolution in real-world environments.

\subsection{Wireless Networks for LLMs}

\begin{figure*}[t]
\centering
\includegraphics[width = 0.9\textwidth]{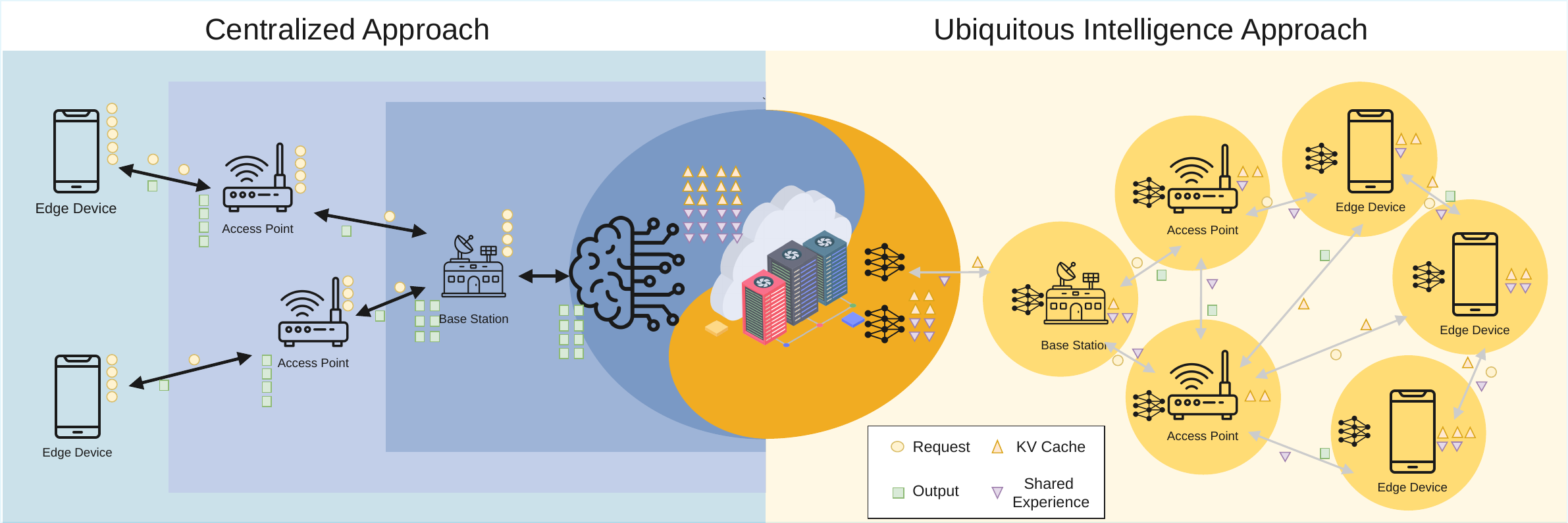}
\caption{Comparison between conventional cloud-centric approach and \emph{ubiquitous intelligence} approach. In the traditional approach, all data and computation are routed through centralized cloud infrastructure, leading to latency, privacy risks, and limited personalization. The \emph{ubiquitous intelligence} framework leverages BSs, access points, and edge devices as active intelligence nodes, enabling local inference, cooperative learning, and D2D knowledge sharing. The distributed structure supports low-latency adoption and enhances resilience and personalization in dynamic wireless environments.}
\label{fig_ubiquitous_intelligence}
\end{figure*}

\subsubsection{Experience Scaling}
% Net4AI 1
While human-generated data remains central to current LLM paradigms, its scalability is limited. In contrast, vast amounts of data from multi-agent systems engaging with diverse environments remain largely untapped. System-wide gathering, processing, and leveraging such interactive experiences represent a critical direction supporting \emph{continuous intelligence ascension} for the next evolutionary era of LLMs scaling.

\subsubsection{Edge-Cloud Network-aided Collaborative Reasoning}
% Net4AI 2
Inference begins at the user terminal, where local computations often take place under constrained resources. Guided by the principle of \emph{system-orchestrated intelligence adaptation}, clusters use low-latency interconnects to map heterogeneous inference tasks to suitable devices, enabling collaborative execution under varying workloads and hardware constraints. This first layer of intelligence leverages local context and device status to make fine-grained task decisions. Such localized reasoning forms the entry point of the system's adaptive intelligence pipeline, balancing response immediacy with offloading potential through wireless networks.

\subsubsection{Distributed Cached Knowledge Management and Propagation}
% Net4AI 2
% split one to 4
D2D communication enables peer-to-peer information exchange, distributing reasoning tasks and intermediate knowledge while reducing reliance on centralized infrastructure~\cite{ioannou2024distributed}. This localized cooperation enhances decentralization and adaptability, supporting the principle of \emph{system-orchestrated intelligence adaptation} across heterogeneous environments. By transferring higher-level reasoning patterns among users, D2D accelerates generalization without full retraining, while distributed knowledge management refines global models and reduces redundancy through intelligent allocation~\cite{pivoto2025comprehensive,hamdi2024network}. Efficiency is further improved through edge caching, predictive prefetching, and opportunistic D2D clustering, enabling robust, low-latency cross-user knowledge management~\cite{liu2024cachegenkvcachecompression}.

\subsubsection{Spectrum-Aware Adaptation for Distributed Learning}
% Net4AI 2
Maintaining performance for resource scheduling in a shared and volatile spectral environment. Spectrum-aware strategies enhance learning stability by embedding real-time channel sensing into the update and synchronization protocols~\cite{nguyenkha2025dtaidedresourcemanagementspectrum}. Devices can regulate their transmission power and update cadence based on interference levels and spectrum availability. These mechanisms play a pivotal role in achieving \emph{system-orchestrated intelligence adaptation} via managing contention and preserving throughput, especially in dense deployments where spectral conditions fluctuate rapidly. Such spectral intelligence contributes directly to the robustness and efficiency of model dissemination.

\subsection{LLMs for Wireless Networks}
\subsubsection{Adaptive Task Offloading Across Edge Devices}
% AI4Net 3
Once local decisions are made, the system dynamically determines whether to retain tasks locally or offload them with experience to more capable edge nodes. Nearby servers or peer devices can assume responsibility for intensive post-inference workloads such as model refinement or skill module execution. Task offloading strategies account for communication quality, processing load, and power availability~\cite{liu2019dynamic} while experience redistribution strategies primarily focus on bandwidth. This adaptive balance between local and distributed computation not only reduces end-to-end latency but also mitigates energy consumption on user devices, ensuring continuity of reasoning under mobility and hardware heterogeneity. Additionally, dynamic allocation of bandwidth and spectrum resources~\cite{sanjalawe2025review} allows distributed learning and inference to remain reliable under fluctuating interference and load conditions. Through continuous adaptation to changing environments, wireless network infrastructures sustain the robustness and responsiveness that realize \emph{permeable semantic networking} at scale.

\subsubsection{Context-Aware Scheduling of Network Resources}
% AI4Net 3
To support seamless task distribution, wireless networks must ensure reliable and efficient delivery of collective LLMs updates. Resource scheduling mechanisms allocate bandwidth, compute cycles, and transmission windows in real time, guided by user behavior, link conditions, and service-level requirements. The system scheduler continuously refines allocations based on updated channel state information and application context, thereby reducing delivery latency and improving system responsiveness~\cite{ornee2023contextawarestatusupdatingwireless}. This layer of context-awareness ensures consistent adaptation and facilitates knowledge sharing across users, even in dynamic and congested wireless environments, supporting \emph{permeable semantic networking}.

\subsubsection{Hierarchical Orchestration Across Network Tiers}
% AI4Net 3
At the core of the wireless intelligence system lies a coordinated architecture that spans multiple layers of the infrastructure. Small cells, macro base stations, and cloud servers collaborate to manage the global distribution of updates and computational resources~\cite{lin2025hierarchicalsplitfederatedlearning} to achieve the goal of \emph{permeable semantic networking}. The orchestration framework determines which components should be processed locally, cached regionally, or distributed globally, based on topology, load conditions, and service priorities. This hierarchical model alleviates backhaul congestion, balances workloads, and ensures scalable deployment of context-specific intelligence. Ultimately, it transforms the wireless network from a passive conduit into an active cognitive substrate.

\subsection{Harnessing Ubiquitous Intelligence for a Greener Future}
Data centers, the backbone of cloud-based LLMs deployment, are rapidly becoming major energy consumers, with U.S. facilities using 4.4\% of national electricity consumption in 2023 and projected to consume 6.7–12\% by 2028 due to escalating AI workloads~\cite{guidi2024environmentalburdenunitedstates}, a trend accelerated by new mega-facilities like OpenAI’s Stargate~\cite{OpenAI2025Stargate}. 
\emph{Ubiquitous intelligence} mitigates this trajectory through redistributing computation to network edges, exploiting underutilized electricity in microgrids with surplus generation and limited storage~\cite{erices2021hairy}. 
Unlike conventional data centers that over-provision backup resources (e.g., batteries and generators) to buffer demand fluctuations but leave them idle during low traffic~\cite{shi2016leveragingenergystorageoptimize}, our framework dynamically routes LLMs' workloads across intelligent edge nodes based on demand and energy availability, thereby harnessing wasted capacity and reducing carbon impact.

\section{Challenges and Open Questions}

Achieving truly \emph{ubiquitous intelligence} over wireless networks requires overcoming several fundamental challenges arising from the interplay between communication, computation, and distributed learning. Key considerations include:

\begin{itemize}
\item \textbf{Scalable experience exchange.} The widespread dissemination of model updates and learned representations generates a substantial communication load. Efficient encoding, transmission prioritization, and adaptive scheduling are needed to ensure a timely and scalable experience sharing across wireless networks, aided by edge nodes.

\item \textbf{Global model consistency.} Maintaining coherence alongside decentralized LLMs remains difficult in dynamic wireless network settings, particularly under intermittent connectivity and asynchronous updates~\cite{chen2023decentralizedlearningwirelessnetworks}. Mechanisms for synchronization, alignment and reconciliation are essential to preserve learning stability and convergence.

\item \textbf{Communication efficiency and latency management.} Exchanging intermediate features or knowledge modules drastically increases pressure on limited bandwidth, while the close integration of inference and communication introduces strict latency constraints. Real-time edge decision-making requires context-aware compression, progressive transmission, and synchronized delay control across computation and communication.

\item \textbf{Security and robustness.} Distributed learning in open wireless environments exposes models to malicious updates, biased feedback, and privacy breaches. Safeguards such as secure aggregation, differential privacy, and anomaly detection are vital for preserving trust and integrity.

\item \textbf{Efficient knowledge representation.} The diversity of experiential data across edge environments naturally gives rise to multi-modal information. Efficient representation of heterogeneous inputs poses significant challenges for unified knowledge integration, efficient retrieval, and downstream usability. Effective solutions must integrate this heterogeneous data in an LLMs-native manner.

\end{itemize}

\section{Conclusion}
\emph{Ubiquitous intelligence} emphasizes the co-evolution of LLMs and wireless networks, where intelligence resides within the wireless infrastructure, and itdynamically coordinating across varied devices to ensure scalable, context-driven, and personalized service delivery. Distributed, context-aware, and adaptive learning across cloud, edge, and device tiers enables models to evolve continuously while meeting pressing demands for low latency, personalization, privacy, and energy efficiency. The intertwined development of wireless networks and LLMs establishes resilient, scalable, and environmentally responsible intelligence that expands through real-world experience. The shift from passive, static LLMs deployments to active, cognitive intelligent systems through wireless networks provides the foundation for continuous-learning capable, dynamic adaptive next-generation AI.

\bibliographystyle{IEEEtran}
\bibliography{Ref}
\end{document}